\documentclass[10pt,journal]{IEEEtran}
\usepackage{amssymb}
\usepackage{graphicx}
\usepackage{picinpar}
\usepackage{booktabs}
\usepackage{float}
\usepackage{amssymb}
\usepackage{url}
\usepackage{cite}
\usepackage{mdwlist}
\usepackage{tikz}
\usepackage{amsfonts}
\usepackage{mathrsfs}
\usepackage{amssymb}
\usepackage{amsmath}
\usepackage{amsthm}
\usepackage{dsfont}
\usepackage{multirow}
\usepackage{color}
\usepackage{graphicx}
\usepackage{subfigure}
\usepackage{subfloat}
\usepackage{epstopdf}
\usepackage{booktabs, tabularx}
\usepackage{array}
\usepackage{slashbox}
\usepackage{algorithmic}
\usepackage[ruled,vlined,linesnumbered]{algorithm2e}

\def\BibTeX{{\rm B\kern-.05em{\sc i\kern-.025em b}\kern-.08em
    T\kern-.1667em\lower.7ex\hbox{E}\kern-.125emX}}
\usepackage{multicol}
\usepackage{microtype}
\usepackage{booktabs} % for professional tables
\usepackage{xcolor}

\usepackage{hyperref}

\ifCLASSINFOpdf

\fi
\begin{document}
\title{CosmoVAE: Variational Autoencoder for\\ CMB Image Inpainting\thanks{Edited: \today}}

\author{Kai Yi, Yi Guo, Yanan Fan, Jan Hamann, Yu Guang Wang$^{*}$

\thanks{K. Yi, Y. Guo, Y. Fan, Y. G. Wang are with the School of Mathematics and Statistics, The University of New South Wales, Sydney, Australia (e-mails:
kai.yi@student.unsw.edu.au; dennis.guo.china@gmail.com;  y.fan@unsw.edu.au; yuguang.wang@unsw.edu.au).}
\thanks{J. Hamann is with the School of Physics, The University of New South Wales, Sydney, Australia (email: jan.hamann@unsw.edu.au).}
\thanks{*~Corresponding author}}
\markboth{}  %
{\MakeLowercase{\textit{et al.}}: }
% The only time the second header will
\maketitle

\begin{abstract}
Cosmic microwave background radiation (CMB) is critical to the understanding of the early universe and precise estimation of cosmological constants. Due to the contamination of thermal dust noise in the galaxy, the CMB map that is an image on the two-dimensional sphere has missing observations, mainly concentrated on the equatorial region. The noise of the CMB map has a significant impact on the estimation precision for cosmological parameters. Inpainting the CMB map can effectively reduce the uncertainty of parametric estimation. In this paper, we propose a deep learning-based variational autoencoder --- CosmoVAE, to restoring the missing observations of the CMB map. The input and output of CosmoVAE are square images. To generate training, validation, and test data sets, we segment the full-sky CMB map into many small images by Cartesian projection. CosmoVAE assigns physical quantities to the parameters of the VAE network by using the angular power spectrum of the Gaussian random field as latent variables. CosmoVAE adopts a new loss function to improve the learning performance of the model, which consists of $\ell_1$ reconstruction loss, Kullback-Leibler divergence between the posterior distribution of encoder network and the prior distribution of latent variables, perceptual loss, and total-variation regularizer. The proposed model achieves state of the art performance for Planck \texttt{Commander} 2018 CMB map inpainting.
\end{abstract}


\IEEEpeerreviewmaketitle

\section{Introduction}

Cosmic Microwave Background (CMB) map details the cooled remnant of the light or electromagnetic radiation caused by the Big Bang in the early stages of the universe,  which one can still observe today \cite{Planck2018I}. The CMB is a valuable resource containing information about how the early universe was formed. It is at a uniform temperature with small fluctuations visible only with high precision telescopes. By measuring and understanding fluctuations, cosmologists can learn the origin of galaxies and explore the basic parameters of the Big Bang theory. The CMB map is produced by map-marker using component separation such as \texttt{Commander} \cite{Er_etal2006,Er_etal2008}, \texttt{NILC} \cite{De_etal2009}, \texttt{SEVEM} \cite{Fe_etal2012} and \texttt{SMICA} \cite{Ca_etal2008}. Due to contamination of thermal noise in the galaxy, the region near the equator of the map (here the equator corresponds to the galaxy) has missing observations for CMB. The left picture of Figure~\ref{fig:cmb_mask} shows the Planck 2018 \texttt{Commander} CMB map, where the noise near the equator is apparent. The right panel of the picture shows the zoomed-in image containing the missing observations at noisy pixels.

\begin{figure}[t]
\centering
\begin{minipage}[t]{\columnwidth}
\centering
\includegraphics[width=\columnwidth]{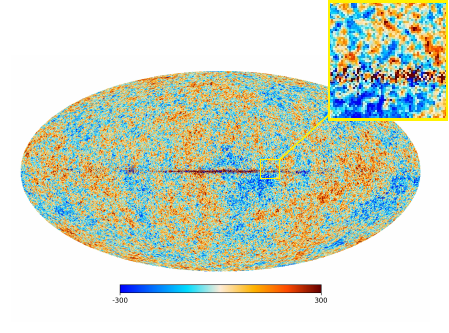}
\end{minipage}
\caption{Planck 2018 \texttt{Commander} CMB Map. The zoomed-in picture is a segmented image used as a test data in the CosmoVAE model, where the dark red indicates the missing pixels to be restored.}\label{fig:cmb_mask}
\end{figure}

Estimating, or inpainting missing observations is a largely unsolved problem in CMB research \cite{cabella2009statistical}. In this work, we propose a new inpainting method for restoring the missing observations  --- CosmoVAE, stemming from a Bayesian deep learning model. CosmoVAE is a variational autoencoder that takes deep convolutional neural networks as encoder and decoder and the whole model as Bayesian inference. The CMB field is a realization of the Gaussian random field on the two-dimensional unit sphere. The CMB map in high resolution is a big data set (for example, in resolution $N_{\rm Side}=2048$, the data is stored at more than 50 million HEALPix points \cite{Gorski_etal2005}).
We show that this Bayes based deep learning model provides an efficient method for inpainting CMB maps using big data of CMB.

\section{Literature Review}
CMB maps can be produced by the following four principles. \texttt{NILC} \cite{De_etal2009} is a linear combination which works in the needlet domain. \texttt{SEVEM} \cite{Fe_etal2012} is a foreground template-cleaning approach that works in the pixel domain to component separation. \texttt{SMICA} \cite{Ca_etal2008} is a non-parametric approach using spherical harmonics as basis. \texttt{Commander} \cite{Er_etal2006,Er_etal2008} is an MCMC based map maker. In \texttt{Commander}, the CMB map is viewed as a realization of Gaussian random field which can be generated by angular power spectrum $C_{\ell}$. The density function of power spectrum $C_{\ell}$ is given by Gibbs sampling and the likelihood of $C_{\ell}$ is estimated by Bayesian inference.
Also, one can use Blackwell-Rao approximation for estimating the angular power spectrum \cite{chu2005cosmological}. The produced CMB map, however, has missing observations. Most mask region is concentrated on the equatorial region due to thermal dust in the galaxy. Inpainting the CMB map is significant to reducing the uncertainty of the estimate for cosmological parameters by CMB data. There have been several methods for CMB map inpainting. For example, Planck consortium \cite{akrami2018planck} uses Gaussian constrained realization to replace the high-foreground regions. Also, one can use Gaussian process regression for CMB image inpainting, which estimates the covariance function for the image pixels and then interpolates the missing pixels \cite{Aghamousa_2017,williams2006gaussian}. Although the traditional statistical method is algorithmically easy to combine different temperature and polarization estimators, they are computationally expensive for the large-sized data set of CMB. According to Gruetjen et al. \cite{gruetjen2017using}, inpainting is an alternative way to construct accurate cut-sky CMB estimators. Gruetjen showed that one could apply inpainting to the problem of unbiased estimation of the power spectrum, which utilizes the linearity of inpainting to construct analytically debiased power spectrum estimates from inpainted maps. 

Recently, with deep learning, inpainting methods have significantly improved reconstruction results by learning semantics from large scale data set. These methods typically use different kinds of convolutional neural networks as mapping functions from masked images to inpainted images end-to-end. Context encoder \cite{pathak2016context} is the first algorithm that uses a deep learning approach to reconstruct masked images. It utilizes the auto-encoder architecture and convolutional neural network with reconstruction and adversarial loss for inpainting. It can achieve surprisingly good performance for restoring an image with a square mask hole. Yang et al. \cite{yang2017high} takes the result from context encoders as input and then propagates the texture information from non-hole regions to fill the hole regions as post-processing. Yu et al. \cite{yu2018generative} proposed an end-to-end image inpainting model with global and local discriminators to ensure the color and texture consistency of generated regions with surroundings. This method has no limitation on the location of mask regions, but the mask shape needs rectangular. As the real CMB mask region is irregular, this method is not suitable. To achieve better inpainting performance for irregular masks, partial convolution \cite{liu2018image} was proposed by Liu et al., where the convolution operation can skip the missing pixels and only use valid pixels. This specified convolution operation can appropriately process with irregular mask and would not lead to artifacts such as color discrepancy and blurriness. With the combination of reconstruction loss, perceptual loss \cite{gatys2015neural}, and total variation loss as penalty term \cite{johnson2016perceptual}, the model achieves state-of-the-art image inpainting results on large data sets such as human faces and landscapes.

 Researchers have proposed many generative probabilistic models based on neural networks in the past decade. Variational autoencoder (VAE) \cite{kingma2013autoencoding} is one of the most popular approaches. With a well-trained VAE model, we can generate various kinds of images by the sampling latent variable with specific distribution (e.g., Gaussian distribution). In many cases, one is interested in training the generative models conditional on the image features such as labels and characteristics of the human face. Sohn et al. \cite{sohn2015learning} proposed conditional variational autoencoder whose input observations modulate the prior on Gaussian latent variables, which then generate the outputs by the decoder. After training, it can specify the output image by controlling the latent variable. Ivanov et al. \cite{ivanov2018variational} modified conditional VAE and proposed variational autoencoder with arbitrary conditioning (VAEAC) model. VAEAC can learn the feature from valid pixels and predict the missing pixels values. Ivanov et al. have used this method for inpainting four different data sets, which achieved state of the art performance.

Our network architecture adopts the auto-encoder architecture, which is widely used in representation learning and image inpainting. We use variational Bayesian approximation to obtain the evidence lower bound (ELBO) of the likelihood of the reconstructed image. The ELBO will be used for our loss function. Besides, we use skip-connection to build a sufficiently deep network and add perceptual loss in the total loss function. We also replace the partial convolution layer \cite{liu2018image} with the vanilla layer, which is more appropriate for CMB image inpainting tasks.

\section{CosmoVAE for CMB}
\begin{figure*}[ht]
\centering
\begin{minipage}{\textwidth}
\centering
\includegraphics[height = 5cm,width=0.9\columnwidth]{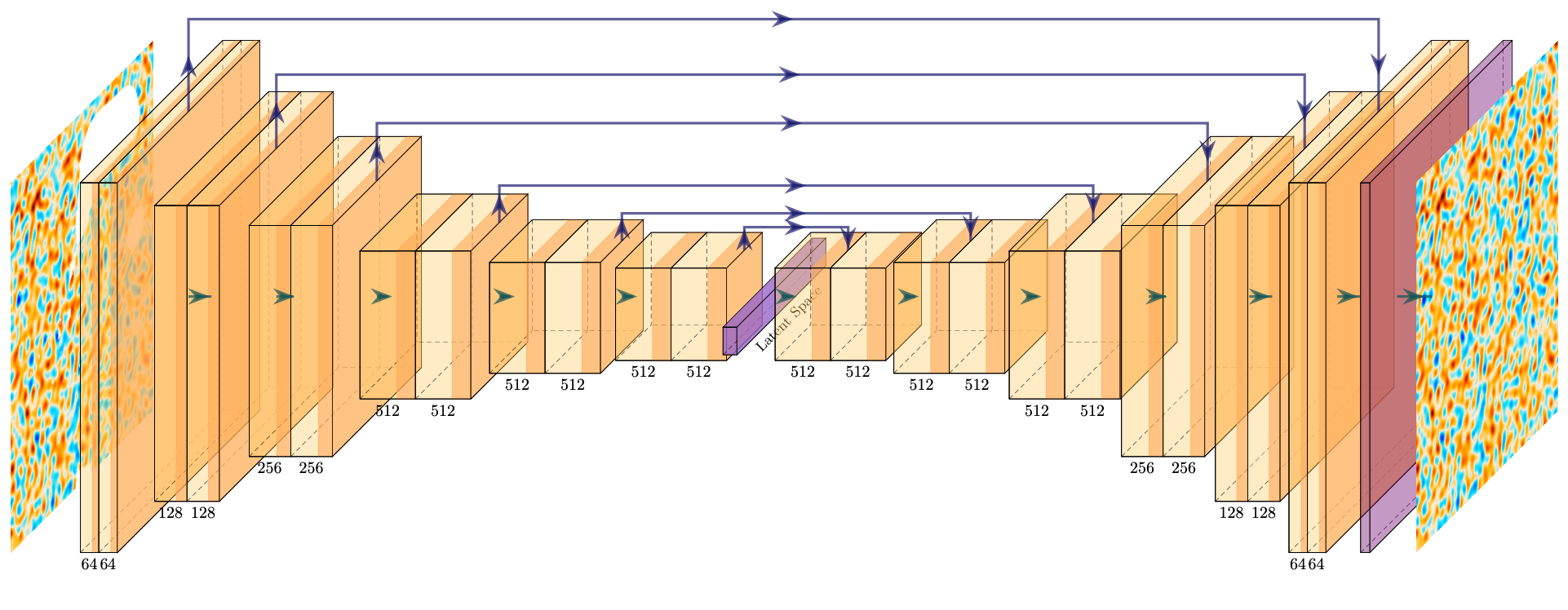}
\end{minipage}
% \hspace{1mm}
\begin{minipage}{0.9\textwidth}
\centering
\caption{Pictorial representation of the Variational Auto-Encoder model. The encoder and decoder, which are connected by channel-wise latent variables, have U-net architecture. The encoder and decoder are deep convolutional neural networks, each with six blocks and three fully-connected layers.}\label{context}
\end{minipage}
\end{figure*}

Our proposed model, as illustrated in Figure~\ref{context}, is based on the variational autoencoders (VAE) \cite{kingma2013autoencoding}, where the encoder and decoder combine the convolutional neural networks (CNNs) and multilayer perceptron (MLP). This modified VAE also uses skip connection between the encoder and decoder, which builds a U-Net-like architecture \cite{liu2018image} in order to guarantee optimal transfer of spatial information from input to the output image. The basic autoencoder compresses the high-dimensional input $x$ (i.e., the segmented image of CMB map) to a low-dimensional latent variable $z$, and then decompresses $z$ back to high-dimensional output $y$, and the input $x$ and output $y$ should be the same.
In the CosmoVAE, the encoder takes the image with a missing region and produces a latent feature representation; the latent features are used by the decoder to produce the missing image. In the training stage, the generated image is compared with the ground truth, where the loss function is composed of the negative variational lower bound, perceptual loss, and a total variation regularizer. A well-trained model can rebuild the mask regions of the CMB map.

\subsection{Statistical Interpretation in VAE}
Let us consider the joint probability distributions of three random variables $((\mathbf{X}, \mathbf{Z},$ $\mathbf{Y})   \in\mathcal{X} \times \mathcal{Z} \times \mathcal{Y})$,
where 
$\mathbf{X}=\left\{\mathbf{x}^{(i)}\right\}_{i=1}^{N}$
is the input masked CMB maps, $\mathbf{Z}$ is the vector of latent variables and $\mathbf{Y}$ is the vector correspsonding to the reconstructed CMB maps. We use neural networks for probabilistic encoder ($Q_{\phi}$) and decoder ($P_{\theta}$). To be precise, the probabilistic encoder is defined as $q_{\phi}(\mathbf{z} | \mathbf{x})$ where $p_{\phi}(z)=\int_{\mathcal{X}} q_{\phi}(z | x) p(x) d x$ for all $z \in \mathcal{Z}$, the $\phi$ denotes the parameters of the neural network. And the probabilistic decoder is given by $p_{\theta}(\mathbf{y} | \mathbf{z})$, with $\theta$  the parameters of the decoder network, and  
\begin{equation*}
    p_{\theta}(y):=\int_{\mathcal{Z}} p_{\theta}(y | z) p(z) d z\quad    \forall y \in \mathcal{Y}.
\end{equation*}
The marginal log-likelihood of output $y$ given by $\log p_{\theta} (y) = \sum_{i=1}^N \log p_{\theta}(y^{(i)})$, for $i=1,\ldots, N$ training samples, can be expressed as variational lower bound which is used as a surrogate objective function where:
%\begin{equation}
\begin{align}\label{eq:ELBO}
    \log p_{\theta}\left(y^{(i)}\right) 
    &\geq \mathbf{E}_{z}\left[\log p_{\theta}\left(y^{(i)} | z\right)\right] 
    \\ \nonumber
    &\qquad - D_{\rm KL}\left(q_{\phi}\left(\cdot | x^{(i)}\right) \| p(\cdot)\right)\\
    &= \mathcal{L}\left(y^{(i)}, x^{(i)}, \theta, \phi\right) \nonumber.
\end{align}
%\end{equation}\label{eq:ELBO}
The variational lower bound consists of negative reconstruction loss and Kullback-Leibler divergence $D_{\rm KL}$ between the approximated posterior $q_{\phi}(z|x)$ of the encoder network and the prior $p(z)$ of the latent variable.

By dual principle, maximizing log-likelihood function $\log p_{\theta}\left(y^{(i)}\right)$ is equivalent to minimizing the negative lower bound $\mathcal{L}\left(y^{(i)}, x^{(i)}, \theta, \phi\right)$ (with respect to the parameters $\theta$ and $\phi$). We thus need to find the gradient of the expectation term. However, this term is intractable, and the variance in the standard Monte Carlo gradient estimators are not computationally efficient. To overcome this, we use the reparameterization trick \cite{kingma2013autoencoding} to find another differentiable estimator, which is computationally friendly.

\subsection{Prior Specification for the  Latent Variables}
The KL-divergence term in the variational lower bound in \eqref{eq:ELBO} can be interpreted as regularization for the parameters $\theta$ and $\phi$, encouraging the approximate posterior $q_{\phi}(z|x)$ to be close to the prior $p(z)$. The posterior distribution $q_{\phi}(z|x)$ can be estimated by probabilistic encoder but the prior distribution remains to be determined.

As CMB is a Gaussian random field and the temperature quantity can be expressed in Fourier series:
\begin{equation*}
	T(\hat{p})=T_{\rm CMB}[1+\Theta(\hat{p})],
\end{equation*}
where $\Theta(\hat{p})$ is the temperature anisotropy in direction $\hat{p}$ which can be expanded with Fourier coefficients $a_{\ell m}$:
\begin{equation*}
	\Theta(\hat{p})=\sum_{\ell =0}^{\infty}\sum_{m=-\ell}^{\ell}a_{\ell m}Y_{\ell m}(\hat{p}).
\end{equation*}

A Gaussian random field is fully determined by the mean and variance. The $a_{\ell m}$ follows a centered Gaussian with variance $C_{\ell}$, where $C_{\ell}$ is the CMB angular power spectrum. Here, we assume that our latent variable ${\bf Z}$ is given by the angular power spectrum $C_{\ell}$. And the generative field is connected with the latent variable by $a_{\ell m } \rightarrow T \rightarrow \mathbf{X}$ and $a_{\ell m } \sim \mathcal{N}(0,C_{\ell})$.
The KL-divergence term can be modified as $D_{\rm KL}\left(q_{\phi}\left(z | x^{(i)}\right) \| \mathcal{N}(0,C_{\ell}) \right)$, the divergence between posterior distribution $q_{\phi}(z|x)$
and the prior distribution. By this, we then assign the physical meaning to the VAE model; that is, the latent variable is the angular power spectrum of the learned field, and the angular power spectrum samples the Fourier coefficients of the generative field.

\subsection{Loss Function}
The overall loss function to train the model is defined as:
\begin{equation*}
	\mathbb{L} = \lambda_{1}\mathbb{L}_{\rm rec} + \lambda_{2}\mathbb{D}_{\rm KL} + \lambda_{3}\mathbb{L}_{\rm perceptual} + \lambda_{4}\mathbb{L}_{\rm TV}.
\end{equation*}
with appropriate weight $\lambda_i$ for each term. Here the weights $\lambda_i$ are hyper-parameters, tuned artificially. The $\lambda_{1}\mathbb{L}_{\rm rec}+\lambda_{2}\mathbb{D}_{\rm KL}$ is the negative variational lower bound which consists of the reconstruction loss and the KL-divergence regularization, the $\mathbb{L}_{\rm perceptual}$ is the perceptual loss\cite{gatys2015neural}, and the $\mathbb{L}_{\rm TV}$ is the total variation loss \cite{johnson2016perceptual}.
\paragraph{Reconstruction Loss}

We use a binary mask $M$ which is 0 for pixel outside the masked region and 1 for pixel inside the masked region. For the network prediction $\hat{y}$ and the ground truth $y$, the reconstruction loss is then
\begin{equation*}
	\mathbb{L}_{\mathrm{rec}} =\frac{1}{N}\left\|(1-M) \odot\left(\hat{y}-y\right)\right\|_{1} +\frac{1}{N}\left\|M \odot\left(\hat{y}-y\right)\right\|_{1},
\end{equation*}
where $N$ denotes normalization constant (where $N = C * H * W$ and $C, H, W$ are the channel size, and the height and width of image).

\paragraph{Regularization}
The regularization term is the KL-divergence between posterior distribution $q_{\phi}(z|x)$
and the prior distribution. The latent variable $a_{\ell m } \sim \mathcal{N}(0,C_{\ell m})$, then, $D_{\rm KL}\left(q_{\phi}\left(z | x^{(i)}\right) \| \mathcal{N}(0,C_{\ell m}) \right)$ can be solved analytically by
\begin{equation*}
	\mathbb{D}_{\rm KL} = \frac{1}{2} \sum_{i=1}^{N}-\log \sigma_{i, 1}^{2}-\log C_{\ell m}^{2}-1+\frac{\sigma_{i, 1}^{2}+\mu_{i, 1}^{2}}{C_{\ell m}^{2}},
\end{equation*}
where $q_{\phi}\left(z | x\right) \sim \mathcal{N}\left(\mu_{1}, \sigma_{1}^2\right)$ is the approximated posterior distribution of the encoder network.

\paragraph{Perceptual Loss}
Perceptual loss is firstly proposed in \cite{gatys2015neural} to preserve image contents in style transfer and is now widely
used for image inpainting. The perceptual loss computes the $\ell_1$ loss of high-level feature maps between the predicted image and ground truth:
\begin{equation*}
    \mathbb{L}_{\text {perceptual}}=\sum_{n=0}^{N-1}\left\|\Psi_{n}^\mathbf{\hat{y}}-\Psi_{n}^\mathbf{y}\right\|_{1},
\end{equation*}
where $\Psi$ is the activation map of the $p$th selected layer which lies in a higher level feature space in ImageNet-pretrained VGG-16 \cite{sundaram2010dense}. We use Pool-1, Pool-2 and Pool-3 layers of VGG-16 for our loss.

\paragraph{Total Variation Loss}
The final term for the loss is the total variation loss as a smoothing penalty:
\begin{align*}
    \mathbb{L}_{\rm TV}
    &= \sum_{(i, j) \in P,\: (i, j+1) \in P}\frac{\left\|\hat{y}^{i, j+1}-\hat{y}^{i, j}\right\|_{1}}{N_{\rm hole}}\\
    &\qquad +\sum_{(i, j) \in P, \:(i+1, j) \in P}\frac{\left\|\hat{y}^{i+1, j}-\hat{y}^{i, j}\right\|_{1}}{N_{\rm hole}},
\end{align*}
where $P$ is the region of 1-pixel dilation of the mask region and $N_{\rm hole}$ is the number of pixels in the mask region \cite{johnson2016perceptual}.

\section{Experimental Results}

\subsection{Generating Training and Test Data Sets}
In the context of CMB map inpainting, preparing training sets has two challenges: the original map is not flat (it resides more naturally on a sphere), and there is only a single CMB map that we can use for training. We can solve the problems by projecting the spherical CMB map to the plane by \emph{Cartesian Projection}.
In order to have a sufficient number of data sets, we segment the whole map with around 50 million pixels into thousands of small images with 400$\times$400 pixels. This is a feasible approach as we assume the CMB field as an isotropic Gaussian random field on the two-dimensional sphere. The resulting small images will be random fields on the squares, which can be assumed independently and identically distributed. We can then treat each small image of the CMB map independently as training data for the deep learning model.

\paragraph{Image data set}
The data of CMB maps are stored at the HEALPix points on the unit sphere \cite{Gorski_etal2005}\footnote{\url{http://healpix.sf.net}}. In the experiments, we use Planck 2018 \texttt{Commander} CMB map with $N_{\rm Side}=2048$ (and  $50,331,648$ points), downloaded from Planck Legacy Archive\footnote{\url{https://pla.esac.esa.int/\#maps}}.
To generate small images cropped from the original map, we project the original map to a flat big 2D image using \emph{Cartesian Projection} of \emph{healpy} (Python) package \cite{zonca2019healpy}. We equally space the points on the projected CMB map, and for each point which then becomes the center of the small image, we take the rectangle whose latitudinal and longitudinal angles ranging from $-5^{\circ}$ to $5^{\circ}$ and $-10^{\circ}$ to $10^{\circ}$ from the center. The spherical CMB full-sky map is then segmented into 4,042 small flat images, each with resolution 400$\times$ 400. The 772 among all small images contain missing regions and will be the test data set, and the remaining clean 3,320 images will be the training set.

\paragraph{Mask data set}
The generative method of mask data set is similar with training data set. We use Planck 2018 Component Separation Inpainting Common mask in Intensity, downloaded from Planck Legacy Archive$^2$, see Figure~\ref{fig:cmb_mask}. The spherical mask map is divided with the same size and same centres as the full-sky CMB map, and segmented into $4,042$ small flat mask images. Each mask image corresponds to a CMB image in training data set, and will be used in masking the missing pixels in training and test. There are $772$ masks having missing pixels, and we randomly select them during the training process.

\begin{figure}[htbp]
\centering
\begin{minipage}[t]{\columnwidth}
\centering
\includegraphics[width=0.95\columnwidth]{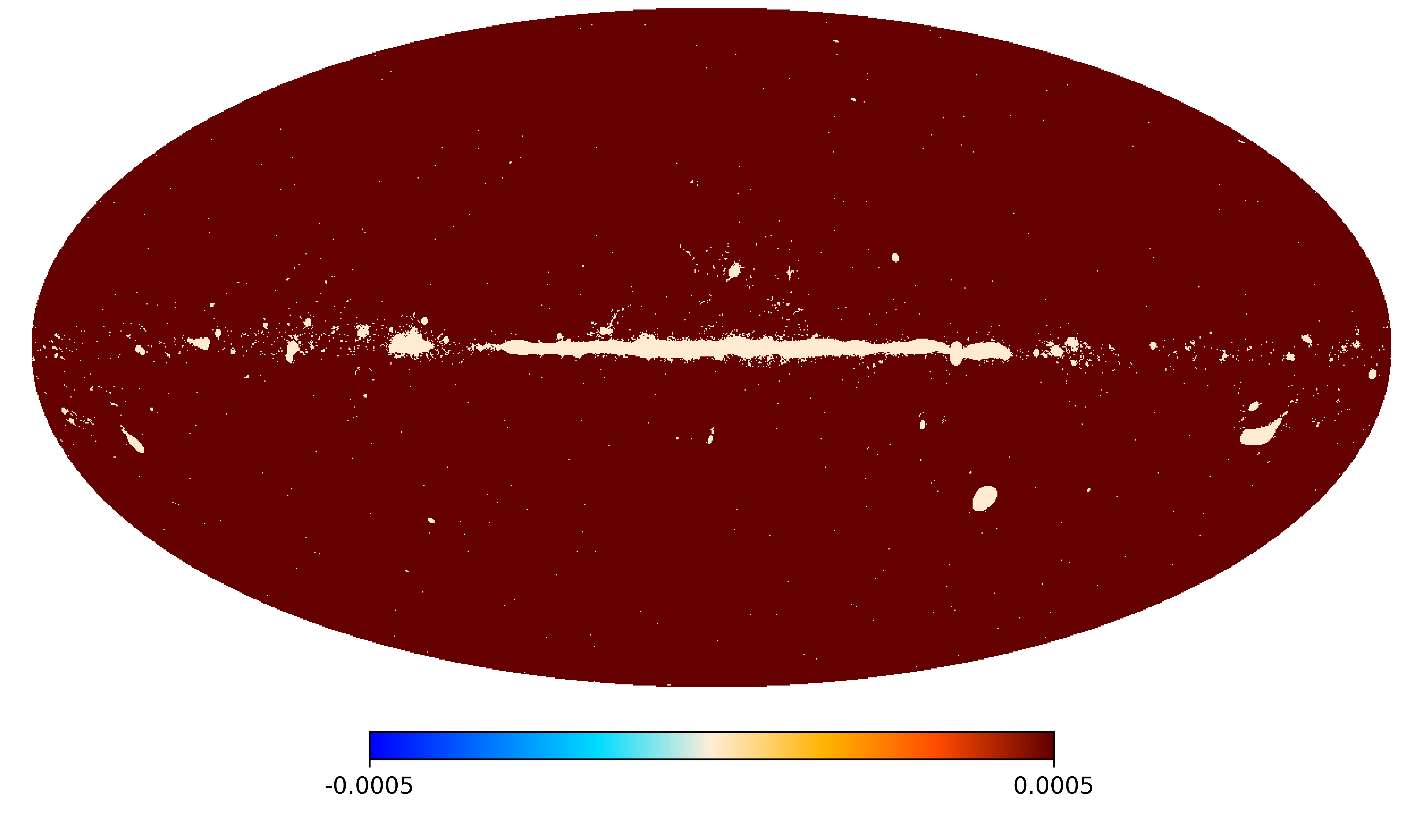}
\end{minipage}
\caption{Planck 2018 Component Separation Inpainting Common mask in Intensity. It is a binary map enciphered with 0 and 1 for clean and noisy pixels. The whole mask map is segmented into 4,042 small images in accord with the small images of CMB full-sky map, each with 400$\times$400 pixels. They are used for masking the full-sky images in training and test.}\label{fig:cmb_mask}
\end{figure}

\subsection{Network Architecture and Training}
Our proposed model is implemented in Keras\footnote{\url{https://github.com/keras-team/keras}}. The network architecture is a U-net-like network. The encoder and decoder have architectures that contain six blocks and three fully connected layers. The encoder network has architecture 64-128-256-512-512-512, and the decoder network architecture is 512-512-512-256-128-64. The latent variable is channel-wise with 2,507 component parameters, which corresponds to angular power spectrum $C_{\ell}$ with $\ell$ up to 2,507. (There are thus the 6,290,064 Fourier coefficients which approximately represent the learned field.) The encoder output samples the latent variable. The whole network is trained using 3,320 400$\times$400 images with batch size four and maximal epoch 1,000. The model is optimized using Adam optimizer with the parameters: learning rate 0.0002 and $\beta_1$ = 0.5. In the training stage, we use the best-fit $\Lambda$CDM CMB TT power spectra from the Planck PR3 baseline\footnote{\url{https://pla.esac.esa.int/\#cosmology}} as $C_{\ell}$.

The experiment is carried out in Google Colab Nvidia Tesla K80 with 2496 CUDA cores, compute 3.7, 12GB GDDR5 VRAM. With no pre-trained weights, it roughly takes 6 hours to achieve convergence. Here the batch size is chosen to adapt to the memory allowed in Colab. If memory is sufficiently large, one can speed up the training by increasing the batch size.

\subsection{Test Results}
\begin{figure}[htbp]
    \centering
    \begin{minipage}{0.3\columnwidth}
    \centering
    \includegraphics[width=\textwidth]{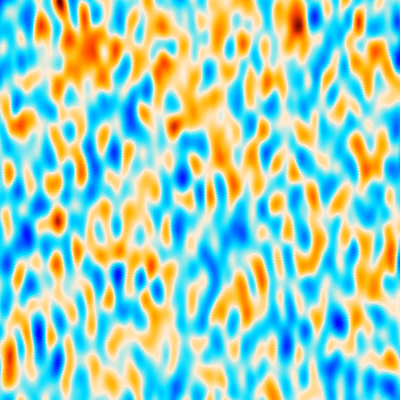}
    {\scriptsize (a) Ground Truth}
    \end{minipage}
    \centering
    \begin{minipage}{0.3\columnwidth}
    \centering
    \includegraphics[width=\textwidth]{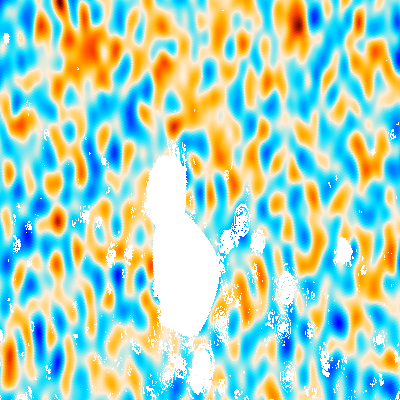}
    {\scriptsize (b) Masked Image}
    \end{minipage}
    \centering
    \begin{minipage}{0.3\columnwidth}
    \centering
    \includegraphics[width=\textwidth]{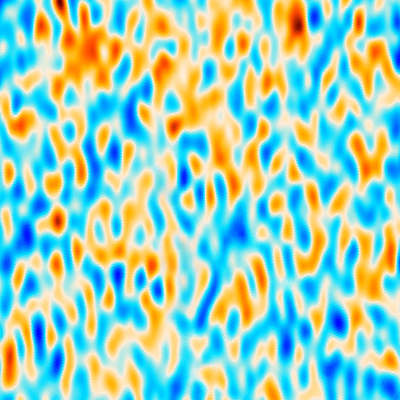}
    {\scriptsize (c) CosmoVAE Predicted}
    \end{minipage}
    \caption{Comparison of predicted results of the proposed models with ground truth. Our models can leverage the surrounding textures and structures and consequently generate lifelike images with no blurriness in the hole area. The mean square error (MSE), mean absolute error (MAE), and peak signal-to-noise ratio (PSNR) are around 0.0055, 0.0134, and 23.989.}
    \label{fig:comparison_test}
\end{figure}

As our test data set consists of masked small CMB images which are not ground truth, we apply the real mask on known small CMB images to evaluate the performance of our model. Consequently, we can check our model's capacity for CMB image inpainting. As shown by Figure~\ref{fig:comparison_test}, 
we compare the training result of CosmoVAE with ground truth image visually and quantitatively. The training indexes: mean square error (MSE), mean absolute error (MAE), and peak signal-to-noise ratio (PSNR) are 0.0055, 0.134 and 23.989 respectively, which reveals the excellent performance of the model. The predicted region by CosmoVAE is apparent as compared with the ground truth. It illustrates that our model can achieve a state of the art performance in training.

\begin{figure}[htbp]
    \centering
\begin{minipage}{\columnwidth}
	\begin{minipage}{0.3\columnwidth}
    \centering
    \includegraphics[width=\columnwidth]{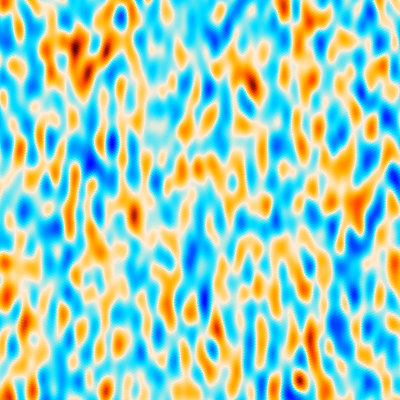}
    \end{minipage}
    \centering
    \begin{minipage}{0.3\columnwidth}
    \centering
    \includegraphics[width=\columnwidth]{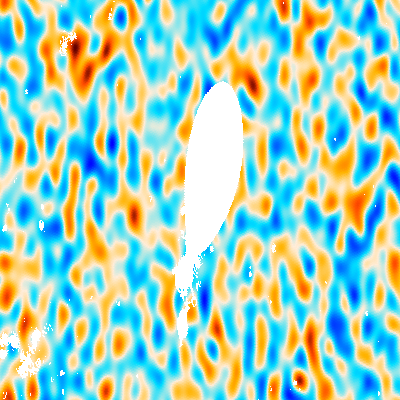}
    \end{minipage}
    \centering
    \begin{minipage}{0.3\columnwidth}
    \centering
    \includegraphics[width=\columnwidth]{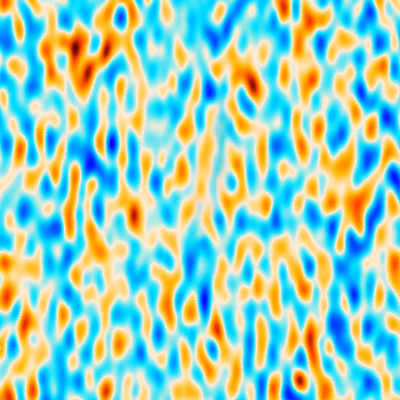}
    \end{minipage}
\end{minipage}\\
    \vspace{6pt}
\begin{minipage}{\columnwidth}
	\begin{minipage}{0.3\columnwidth}
    \centering
    \includegraphics[width=\textwidth]{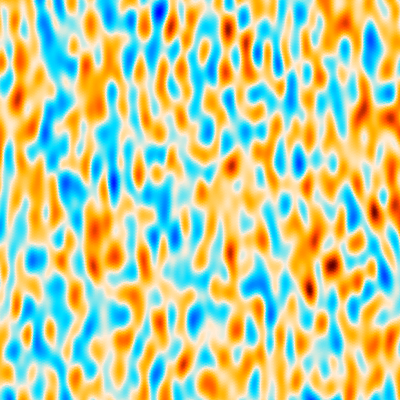}
    \end{minipage}
    \centering
    \begin{minipage}{0.3\columnwidth}
    \centering
    \includegraphics[width=\textwidth]{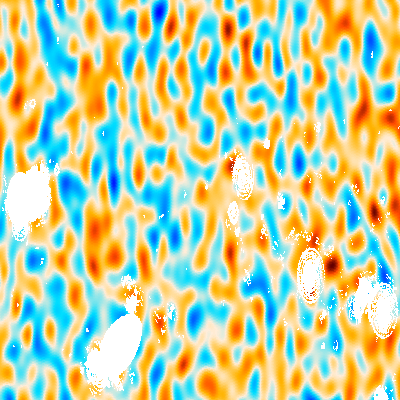}
    \end{minipage}
    \centering
    \begin{minipage}{0.3\columnwidth}
    \centering
    \includegraphics[width=\textwidth]{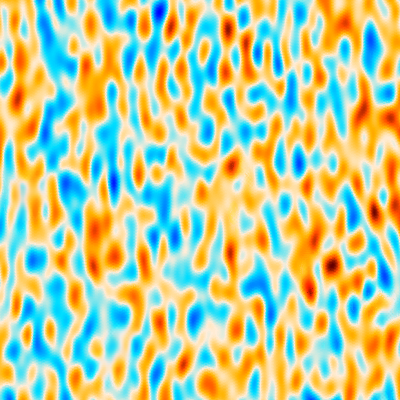}
    \end{minipage} 
\end{minipage}\\
    \vspace{6pt} 
\begin{minipage}{\columnwidth}
	\begin{minipage}{0.3\columnwidth}
    \centering
    \includegraphics[width=\textwidth]{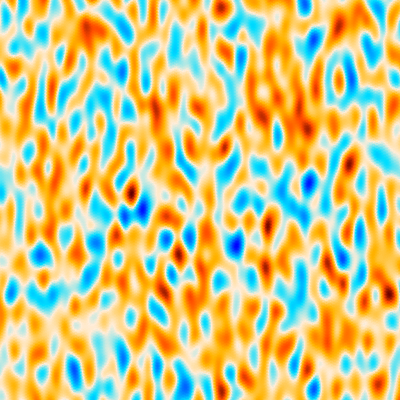}
    \end{minipage}
    \centering
    \begin{minipage}{0.3\columnwidth}
    \centering
    \includegraphics[width=\textwidth]{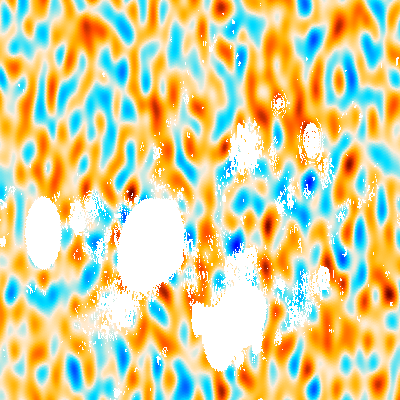}
    \end{minipage}
    \centering
    \begin{minipage}{0.3\columnwidth}
    \centering
    \includegraphics[width=\textwidth]{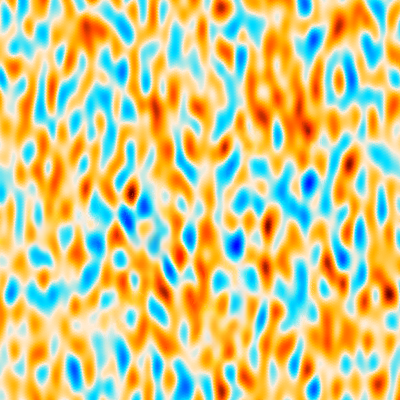}
    \end{minipage}
\end{minipage}\\  
	\vspace{6pt} 
\begin{minipage}{\columnwidth}
	\begin{minipage}{0.3\columnwidth}
    \centering
    \includegraphics[width=\textwidth]{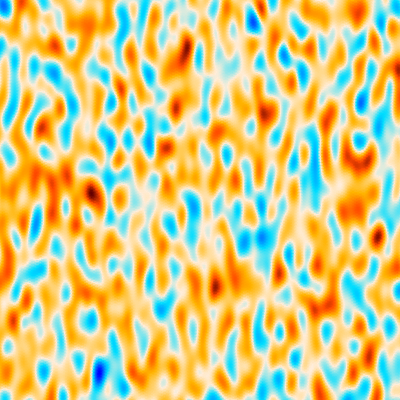}
    \end{minipage}
    \centering
    \begin{minipage}{0.3\columnwidth}
    \centering
    \includegraphics[width=\textwidth]{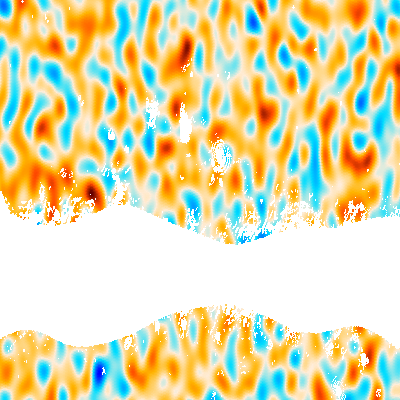}
    \end{minipage}
    \centering
    \begin{minipage}{0.3\columnwidth}
    \centering
    \includegraphics[width=\textwidth]{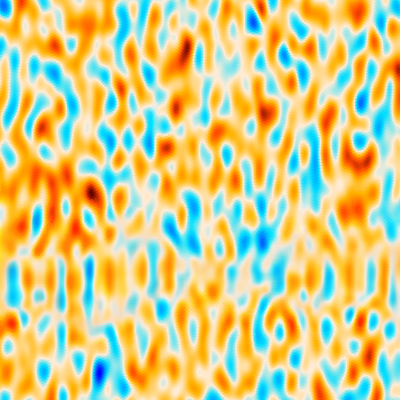}
    \end{minipage}
\end{minipage}\\   
	\vspace{6pt}
\begin{minipage}{\columnwidth}
	\begin{minipage}{0.3\columnwidth}
    \centering
    \includegraphics[width=\textwidth]{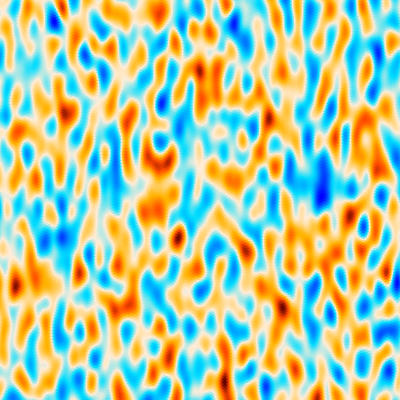}
    \end{minipage}
    \centering
    \begin{minipage}{0.3\columnwidth}
    \centering
    \includegraphics[width=\textwidth]{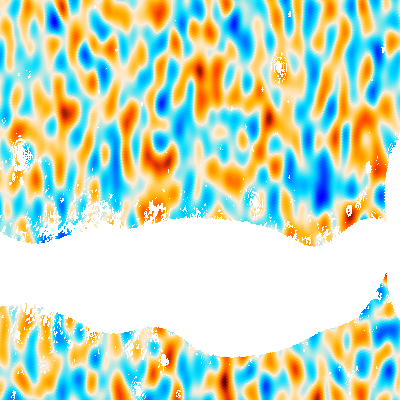}
    \end{minipage}
    \centering
    \begin{minipage}{0.3\columnwidth}
    \centering
    \includegraphics[width=\textwidth]{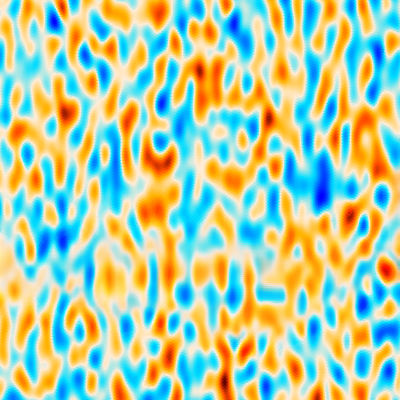}
	\end{minipage}
\end{minipage}  
\vspace{6pt}
\begin{minipage}{\columnwidth}
\caption{Comparison of the results of our CosmoVAE model and Planck 2018 \texttt{Commander}. The left image is Planck 2018 \texttt{Commander} results. The middle is the contaminated image after cropping. The right image is the inpainted image. The CosmoVAE has good performance in predicting the missing observations in various mask regions.}\label{fig:train1}
\end{minipage}
\end{figure}

When having trained the CosmoVAE, we use it to predict the missing pixels of each small CMB image in the test data set. Figure~\ref{fig:train1} shows five examples of the predicted results by CosmoVAE. We compare our inpainted results with Planck 2018 results \cite{akrami2018planck}. The left-most plot shows the inpainted CMB image of Planck 2018 results \cite{akrami2018planck}. The second column shows the original (un-inpainted) CMB image with an irregular missing region, which is the actual input of the trained CosmoVAE. The third column panel shows the corresponding predicted images, where the network restores the missing region. As we can observe, the trained CosmoVAE can inpaint CMB images with irregular mask regions, even if the mask area is big, and there are multiple mask holes. The predicted results are evident as compared with the Planck 2018 results. CosmoVAE thus provides a useful inpainting model for the CMB map.

\subsection{Uncertainty Quantification}
One can interpret the CosmoVAE as a probability model in a similar way to that of AEVB \cite{kingma2013autoencoding}. More concretely, denote the encoder neural networks model mapping input $x$ to a stochastic latent variable $z$ (see Figure \ref{context}) by the conditional probability $p_{\theta}(z|x)$, where $\theta$ denotes the parameters of the encoder network.
Similarly, denote the decoder neural networks model mapping $z$ to the output $x$, as $p_{\phi}(x|z)$ and $\phi$ denote the weights and biases of the decoder network.

In the Bayesian inferential context, $p(z|x)$ is the posterior distribution obtained from the prior and likelihood combination where $p(z|x) \propto p_{\phi}(x|z)p(z)$, and $p_{\phi}(x|z)$ is the likelihood (the "generative" model or, the decoder), and $p(z)$ is the prior.
In the variational inference framework, a distribution $q_{\lambda}(z|x)$ can be used to approximate this intractable posterior $p(z|x)$. 

If we take the latent variable $z$ as a $\mathcal{N}(0,1)$ variate and the relationship between $x$ and $z$ is given by the encoder network with parameter $\theta$. Then taking $q_{\lambda}(z|x)$ to be a Normal distribution with parameters $\lambda=(\mu(\theta),\sigma(\theta))$, minimising the Kullback-Leibler divergence between $q_{\lambda}(z|x)$ and the true posterior $p(z|x)$ is now the same as minimising the loss functions with respect to $\theta$ and $\phi$, where different loss functions $\mathbb{L}(x)$ broadly correspond to the noise distribution we assume for $x$. 
In the CosmoVAE, the parameters $\theta$ depend on the input $x_1$ and $\phi$ depend on $x_2$, where the components of $x=(x_1, x_2)\equiv ((1-M)\odot x, M\odot x)$ are respectively the pixels outside and inside the masked regions.

Having a posterior distribution allows us to obtain uncertainty estimation. To do this, consider the posterior predictive distribution of $x_2$, where $$p(x_2|x_1)= \int p(x_2|z) p(z|x_1)dz, $$  
where $p(z|x_1) = q_{\lambda}(z|x)$ is simply the posterior distribution obtained from the variational approximation, using training data $x$, and  $p(x_2|z)$ is the generative model, and when a discriminator model is added to $p(x_2|z)$, we can sample $x_2$ and retain images for which the discriminator has computed as true. The variability in the pixels of the "true" images provides us with the uncertainty measure.

\begin{figure}
    \centering
    \begin{minipage}{0.24\columnwidth}
    \centering
    \includegraphics[width=\textwidth]{1718Planck.png}
    \end{minipage}
    \centering
    \begin{minipage}{0.24\columnwidth}
    \centering
    \includegraphics[width=\textwidth]{1718Masked.png}
    \end{minipage}
    \centering
    \begin{minipage}{0.24\columnwidth}
    \centering
    \includegraphics[width=\textwidth]{1718CosmoVAE.png}
    \end{minipage}
    \centering
    \begin{minipage}{0.24\columnwidth}
    \centering
    \includegraphics[width=\textwidth]{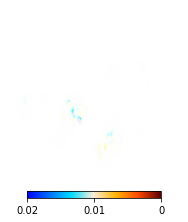}
    \end{minipage}
        \centering
    \begin{minipage}{0.24\columnwidth}
    \centering
    \includegraphics[width=\textwidth]{1738Planck.png}
    \end{minipage}
    \centering
    \begin{minipage}{0.24\columnwidth}
    \centering
    \includegraphics[width=\textwidth]{1738Masked.png}
    \end{minipage}
    \centering
    \begin{minipage}{0.24\columnwidth}
    \centering
    \includegraphics[width=\textwidth]{1738CosmoVAE.png}
    \end{minipage}
    \centering
    \begin{minipage}{0.24\columnwidth}
    \centering
    \includegraphics[width=\textwidth]{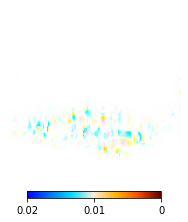}
\end{minipage}
\begin{minipage}{\columnwidth}
    \caption{A closer look at the test result for two images of the Planck 2018 \texttt{Commander} CMB map. The left column plots the inpainted image from Planck 2018 results. The second column plots the original image with the irregular missing region. The third column plot shows the predicted image by CosmoVAE. The right-most column plots the standard deviation of the test outputs for the same sample using $100$ trained models.}\label{fig:test1}
    \end{minipage}
\end{figure}

Figure~\ref{fig:test1} shows the uncertainty in the CosmoVAE predictor. The second column is the CMB image with the missing region masked. The fourth column shows the standard deviation of the inpainted images of the trained-CosmoVAE at each missing pixel over 100 different realizations of the latent variable. Compared with the Planck 2018 result in the first column for the same image segment from the full-sky CMB map, the CosmoVAE inpainted image has the same image quality. The Std Dev. for each image is very small, and the location where the Std Dev. is significant in the square image is a fractional part of the mask region. Thus, the uncertainty of CosmoVAE is controllable and has little effect on the predicted image.

%\subsection{Comparison with Gaussian Process Model}
%The Planck 2018 results as shown in Figure~\ref{fig:comparison_test} is achieved by the 

\section{Conclusion and Future Plan}
Statistical challenges to processing the CMB data is one of the biggest challenges in the analysis of CMB data. In this work, we reconstruct the cosmic microwave background radiation (CMB) map by using a modified variational autoencoder (VAE) as our baseline model. We cut the full-sky CMB map into many small images in order to generate our image and mask datasets, and then in training to inpaint the hole area with arbitrary shape mask. To enhance the performance, we combine our neural network with the angular power spectrum, which can generate the Fourier coefficients of the Gaussian random field. Also, we modify the original VAE loss function by adding in the perceptual loss and the total-variation regularizer. This new VAE model assigns cosmological meaning to the parameters of the network and thus achieves a state of the art performance for CMB map inpainting.

To better complete image inpainting task and for cosmology study, one needs to reconstruct the full-sky CMB map from all small inpainted CMB images. We can use the inpainted full-sky CMB map to estimate cosmological parameters such as the angular power spectrum $C_{\ell}$, which can be computed directly by the healpy package. The inpainting of the CMB map will help reduce the uncertainty in the parametric estimation. 
By Olivier et al. \cite{tolstikhin2018wasserstein}, any method which is based on the marginal log-likelihood, including VAE, necessarily leads to the blurriness of output due to the gap between true negative log-likelihood and the upper bound (ELBO). 
We can further modify our loss function to improve the quality of reconstructed images by replacing the KL-divergence with GAN or WGAN (more stable) as regularizer. Our model can also be a baseline model for the reconstruction of other Gaussian random fields (besides the CMB field). We will probe these problems in our future work. 

\subsubsection*{Acknowledgments}
Some of the results in this paper have been derived using the HEALPix \cite{Gorski_etal2005} package.

\ifCLASSOPTIONcaptionsoff
  \newpage
\fi

%\ifCLASSOPTIONcaptionsoff
%  \newpage
%\fi
%
%% trigger a \newpage just before the given reference
%% number - used to balance the columns on the last page
%% adjust value as needed - may need to be readjusted if
%% the document is modified later


%\bibliographystyle{IEEEtran}
\bibliographystyle{IEEEtran}
\bibliography{cosmo}

%%%%%%%%%%%%%%%%%%%%%%%%%%%%%%%%%%%%%%%%%%%%%%%%%%%%%%%%%%%%%%%%%%%%%%%%%%%%%%%%%%%%%%%%%%

%\newpage

\end{document}